\documentclass[aps,prb,reprint,superscriptaddress]{revtex4-1}

\usepackage{graphicx}
\usepackage{color}
\usepackage{amsfonts}
\usepackage{amsmath}
\usepackage{bbm}
\usepackage[caption=false]{subfig}

\renewcommand{\vec}{\mathbf}

\begin{document}

\title{Theory of scanning gate microscopy imaging \\ of the supercurrent distribution in a planar Josephson junction}

\author{K. Kaperek}
\affiliation{AGH University of Science and Technology, Faculty of Physics and Applied Computer Science, al. A. Mickiewicza 30, 30-059 Krakow, Poland}

\author{S. Heun}
\affiliation{NEST, Istituto Nanoscienze-CNR and Scuola Normale Superiore, Piazza San Silvestro 12, 56127 Pisa, Italy}

\author{M. Carrega}
\affiliation{CNR-SPIN, Via Dodecaneso 33, 16146, Genova, Italy}

\author{P. Wójcik}
\affiliation{AGH University of Science and Technology, Faculty of Physics and Applied Computer Science, al. A. Mickiewicza 30, 30-059 Krakow, Poland}

\author{M. P. Nowak}
\email{mpnowak@agh.edu.pl}
\affiliation{AGH University of Science and Technology, Academic Centre for Materials and Nanotechnology, al. A. Mickiewicza 30, 30-059 Krakow, Poland}

\date{\today}

\begin{abstract}
We theoretically investigate the mapping of the supercurrent distribution in a planar superconductor-normal-superconductor junction in the presence of a perpendicular magnetic field via the scanning gate microscopy technique. We find that the distribution of counter-propagating supercurrents aligned in Josephson vortices can be mapped by the change of the critical current induced by the tip of the scanning probe, if the flux in the junction is set close to maxima of the Fraunhofer pattern. Instead, when the magnetic field drives the junction to a supercurrent minimum in the Fraunhofer pattern, the superconducting phase adapts, and the tip always increases the supercurrent. The perpendicular magnetic field leads to the formation of Josephson vortices, whose extension for highly transparent junctions depends on the current circulation direction. We show that this leads to an asymmetric supercurrent distribution in the junction and that this can be revealed by scanning gate microscopy. We explain our findings on the basis of numerical calculations for both short- and long-junction limits and provide a phenomenological model for the observed phenomena.
\end{abstract}

\maketitle

\section{Introduction}

The supercurrent in superconductor-normal-superconductor (SNS) Josephson junctions is carried by quasiparticles forming Andreev bound states (ABS) in the normal region. Upon application of a perpendicular magnetic field in SNS junctions for which the self field of the supercurrent can be neglected \cite{Barone1982, gross2016applied, PhysRevB.94.094514}, the vector potential of the field induces a spatial gradient of the superconducting phase difference. This in turn introduces a spatial variation of the supercurrent, leading to formation of negative and positive current regions, and ultimately to the formation of supercurrent vortices. As such, vortices correspond to a complete loops of the current and carry no net transport of supercurrent. They are refereed to as Josephson vortices \cite{tinkham2004introduction}. An increase in magnetic field leads to the formation of subsequent vortices \cite{PhysRevLett.11.200}, and accordingly the critical current of the junction exhibits a Fraunhofer oscillation pattern.

The experimental confirmation of the formation of Josephson vortices was obtained by scanning tunneling microscopy \cite{Roditchev2015}, and their manipulation was performed by magnetic force microscopy \cite{Dremov2019}. Moreover, the change of the magnetic interference pattern from Fraunhofer to SQUID-like when the density of states is tuned from bulk to edge states in quasi-ballistic junction \cite{PhysRevB.102.165407} was used for {\it indirect} determination of the supercurrent distribution \cite{Hart2014, Pribiag2015, Allen2016, doi:10.1021/acs.nanolett.0c00025, Guiducci2019}.

Here we theoretically explore the possibility of a {\it direct} visualization of the supercurrent distribution and the associated formation of Josephson vortices by a combination of critical current measurements and the scanning gate microscopy (SGM) technique. Our idea exploits the fact that in novel planar SNS junctions \cite{PhysRevB.93.155402, PhysRevApplied.7.034029, Fornieri2019, Guiducci2019, Guiducci2019a, PhysRevLett.120.047702, Ren2019, Telesio2021, Ke2019, Moehle2021, PhysRevB.99.245302, PhysRevResearch.1.032031, salimian2021gatecontrolled, PhysRevB.98.245418, PhysRevB.93.214502} the normal semiconducting regions between the superconductors remains exposed, which opens the possibility for the application of the SGM technique. The latter technique has been successfully used for over two decades for imaging of normal current flow.

Indeed, in normal systems SGM is a widely used technique which relies on measuring the conductance changes when a biased atomic force microscope tip scans over the device \cite{doi:10.1126/science.1069923}, inducing a repulsive potential in the two-dimensional electron gas (2DEG) of the sample and affecting the trajectories of the propagating electrons. SGM allows for the demonstration of, e.g., magnetic focusing \cite{Aidala2007} or branched electron flow \cite{PhysRevLett.105.166802, PhysRevB.88.035406, PhysRevB.90.035301} in heterostructures \cite{Topinka2001, PARADISO20101038, PhysRevLett.94.126801, PhysRevB.80.041303} and monolayers \cite{NEUBECK20121002, PhysRevB.83.115441, PhysRevB.96.165310, PhysRevB.87.085446, prokop_gut_nowak_2020}. 

Here we theoretically predict that the SGM technique enables spatial imaging of the supercurrent in a Josephson junction realized in a ballistic 2DEG and allows the identification of positive and negative supercurrent streams that contribute to the formation of Josephson vortices. 
However, in contrast to non-superconducting devices, the transport properties and the supercurrent distribution in a SNS junction are controlled by the superconducting phase difference. We show that SGM allows for the visualization of the supercurrent distribution by measuring the critical current of the junction provided that an external transverse magnetic field is set near the Fraunhofer maxima, i.e., when there is an even number of Josephson vortices and the phase is uniquely defined despite changes of SGM tip position. 
Moreover, we show that the supercurrent vortices in a ballistic junction are inherently nonsymmetric, which in turn results in the asymmetry of the supercurrent flowing across the junction, which can be revealed by the SGM technique.
We test our findings in an experimentally viable regime of short and long junctions, paving the way to study the Josephson vortex structure and supercurrent distribution in a variety of planar SNS devices, for which a standard Fraunhofer pattern is currently observed. Furthermore, the proposed method of supercurrent visualization can be applied to systems such as common 2DEGs where scanning tunneling microscopy is not possible due to surface oxidization \cite{Chen1993,Voigtlaender2015}. 

The outline of this paper is as follows. We first introduce the system in Sec.~II. In Sec.~III, we present the results for the short-junction regime. Phenomenological analysis of supercurrent probing is presented in Sec.~III.~A, while in Sec.~III.~B we explain the asymmetry of the Josephson vortices. Finally, in Sec.~IV, we test the robustness of our results beyond the short-junction limit and summarize our results in Sec.~V.

\section{The investigated system}

\begin{figure}[ht!]
\center
\includegraphics[width = 8cm]{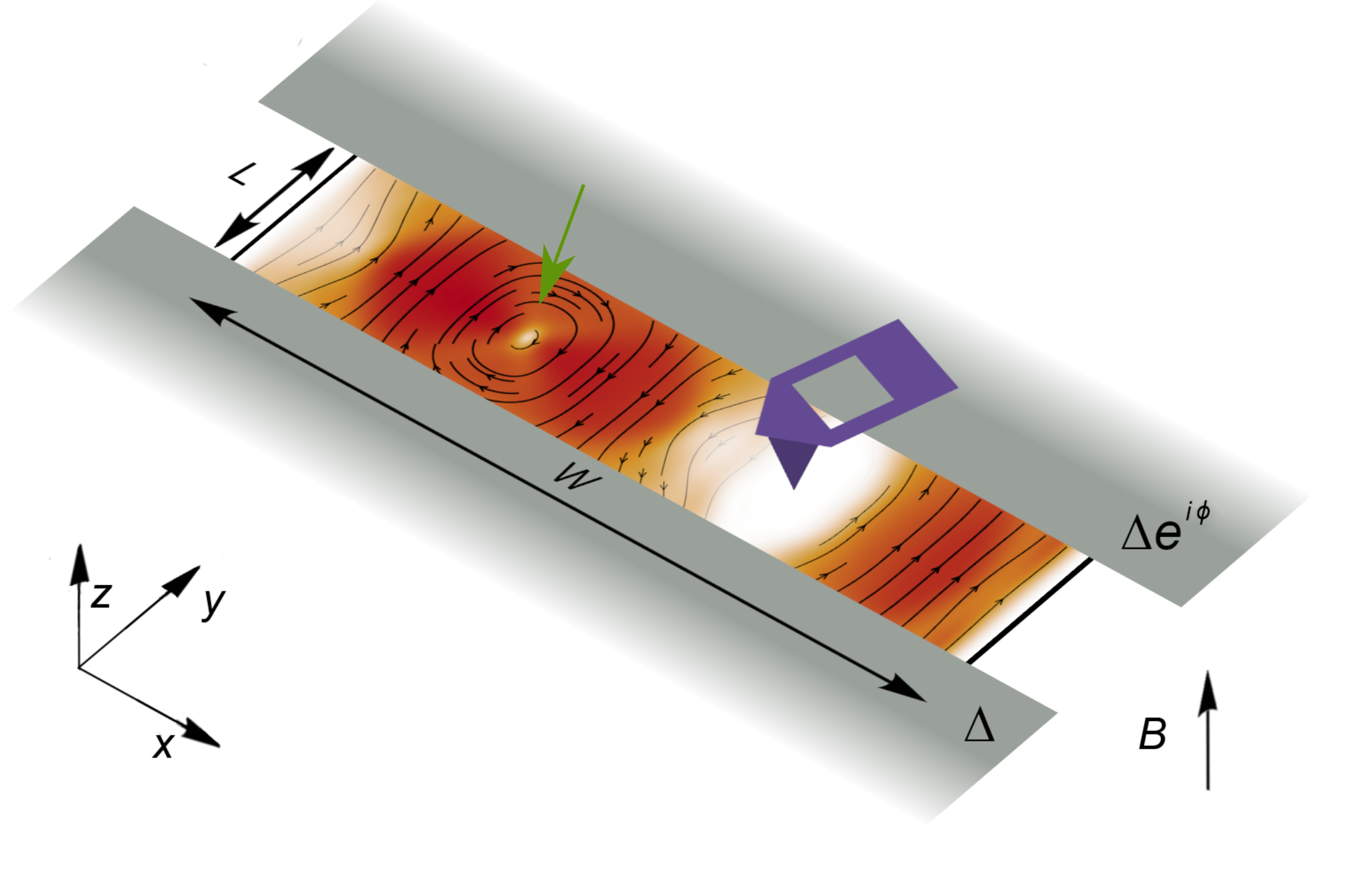}
\caption{Illustration of the considered system. The grey color depicts the superconducting contacts. Between them there is the normal region whose surface is scanned by the SGM tip (purple). With the orange stream-plot we show an exemplary supercurrent distribution with a Josephson vortex denoted with the green arrow.}
\label{system}
\end{figure}

We study a planar SNS junction composed of two superconducting leads connected through a semiconducting region in the ballistic regime, in the presence of a magnetic field oriented perpendicular to the junction plane---see Fig.~\ref{system}. The junction dimensions are defined by the width ($W$) and the length ($L$) of the semiconducting region. In experimental realizations, such planar junctions have been defined in InAs \cite{PhysRevLett.75.3533, PhysRevB.71.174502, PhysRevB.93.155402, PhysRevApplied.7.034029, Fornieri2019, Guiducci2019, Guiducci2019a, PhysRevLett.120.047702}, HgTe \cite{Ren2019}, bP \cite{Telesio2021}, graphene \cite{GrapheneJJs, Calado2015, Lee2018, BenShalom2016}, InSb \cite{Ke2019} or InAsSb \cite{Mayer2020, Moehle2021} quantum wells, and InSb nanoflakes \cite{PhysRevB.99.245302, PhysRevResearch.1.032031, salimian2021gatecontrolled} interfaced with aluminum or niobium superconducting contacts. The supercurrent (visualized in orange in Fig.~\ref{system}) carried by the quasiparticles in the semiconductor is disturbed by the repelling potential of a negatively charged SGM tip (purple in Fig.~\ref{system}) scanning above the surface of the semiconductor.

\section{Short-junction regime}

We start with calculations in the short-junction approximation, i.e., in the regime where the length of the normal scattering region is much shorter than the superconducting coherence length, $L \ll \xi = \hbar v_F /  \Delta$, with $v_F$ the Fermi velocity in the semiconductor and $\Delta$ the superconducting gap. This corresponds to the case where the dwell time of quasiparticles inside the normal region $\tau_{\mathrm{dw}}$ is much smaller than the time $\hbar/\Delta$ it spends inside the superconductor.\cite{shortjunctionnote}

Between the superconducting contacts, the Andreev-reflected electrons and holes create a set of bound states, whose energies can be determined \cite{PhysRevLett.67.3836} from a matching condition $S_A(E)S_N(E) \Psi_{\mathrm{in}} = \Psi_{\mathrm{in}}$ where $\Psi_{\mathrm{in}}=(\Psi_e, \Psi_h)$ are the complex amplitudes of the electron and hole waves incident on the junction defined in the basis of normal region scattering modes. $S_A(E)$ describes an Andreev reflection process at the NS interface
\begin{equation}
S_{A}(E)=\zeta(E)\left(\begin{array}{cc}
0 & r_A^* \\
r_A & 0 \\
\end{array} \right),
\label{SA_array}
\end{equation}
with $\zeta(E) = \sqrt{1-E^2/\Delta^2}+iE/\Delta$.
Taking the outgoing modes as time-reversed partners of the incoming states and assuming that Andreev reflection does not mix the scattering modes, we can write
\begin{equation}
r_{A}=\left(\begin{array}{cc}
i\mathbf{1} & 0 \\
0 & ie^{-i\phi}\mathbf{1} \\
\end{array} \right),
\label{r_array}
\end{equation}
with $\phi$ the superconducting phase difference and $\mathbf{1}$ the identity matrix spanning on the basis of the scattering modes.

The scattering properties for electrons and holes in the normal part are captured in the block-diagonal matrix
\begin{equation}
S_{N}(E)=\left( \begin{array}{cc}
S(E) & 0 \\
0 & S^*(-E) \\
\end{array} \right),
\label{SN_array}
\end{equation}
where $S(E)$ ($S^*(-E)$) corresponds to the electron (hole) scattering block. Taking advantage of the short-junction approximation, i.e. $S(E) = S(E=0) \equiv s$, we arrive at the eigenproblem \cite{PhysRevB.90.155450}
\begin{equation}
\left(\begin{array}{cc}
s^\dagger & 0 \\
0 & s^T \\
\end{array} \right)
\left(\begin{array}{cc}
0 & r^*_{A} \\
r_{A} & 0 \\
\end{array} \right) \Psi_{\mathrm{in}}  = \zeta(E) \, \Psi_{\mathrm{in}},
\label{ABS_eigenequation}
\end{equation}
whose solution yields the set of ABS eigenenergies and wave functions.

The supercurrent is obtained from the positive energy ABS,\cite{PhysRevLett.67.3836}
\begin{equation}\label{tot_current}
I=-\frac{e}{\hbar}\sum_{E_i>0}\tanh\left(\frac{E_i}{2k_{\rm B}T}\right)\frac{d E_i}{d \phi},
\end{equation}
and consequently, the critical current is $I_c = \max_\phi(I)$. 

The scattering matrix is obtained for the normal region of the junction described by the Hamiltonian
\begin{equation}
H_N = \left( \frac{\hbar^2 \mathbf{k}^2}{2m^*} - \mu + V_{\mathrm{tip}}(x,y)\right)\sigma_0 + \alpha(\sigma_x k_y - \sigma_y k_x).
\label{HN}
\end{equation}
The last term corresponds to Rashba spin-orbit coupling, $\mu$ is the chemical potential, and $\sigma_0$ is an identity matrix acting on the spin degree of freedom.

The impact of the tip positioned at ($x_{\mathrm{tip}}, y_{\mathrm{tip}}$) on the scattering of quasiparticles is included through the $V_{\mathrm{tip}}$ term in the Hamiltonian Eq.~(\ref{HN}) that induces a Lorentzian potential island \cite{PhysRevB.84.075336} in the 2DEG
\begin{equation}
V_{\mathrm{tip}}(x,y) = \frac{V_0}{1+\frac{(x-x_{\mathrm{tip}})^2 +(y-y_{\mathrm{tip}})^2}{d^2}},
\end{equation}
where we take  $V_0 = 100$~meV as the effective potential magnitude (corresponding to a negative voltage on the tip) and $d = 50$~nm its width \cite{PhysRevB.84.075336}.

We discretize the Hamiltonian Eq.~(\ref{HN}) which results in a set of onsite and hopping elements that are used to describe the system on a square computational mesh. We use a lattice constant $\delta x = \delta y = 10$~nm. The orbital effects of the magnetic field are included via Peierls substitution of the hopping elements $t_{nm} \rightarrow t_{nm}\exp\left[-ie/\hbar\int\mathbf{A} d\mathbf{l}\right]$ where the integral is taken over the hopping direction. We adopt the vector potential in the Landau gauge $\mathbf{A} = (0,xB,0)$. For the sake of simplicity, we neglect the Zeeman interaction at this stage, whose impact will be included explicitly in long-junction calculations in the following. The electron scattering matrix $S(E=0)$ of the normal region is obtained using the Kwant package \cite{Groth_2014}. In the scattering matrix calculation, to take into account the electrons incident on the normal region, we consider normal, semi-infinite leads and assume zero magnetic field within them to account for the screening effect.

For definiteness, we adopt the material parameters corresponding to InSb \cite{doi:10.1063/1.1368156}, i.e., effective mass $m^* = 0.014m$,  with $m$ the bare electron mass. We assume a wide junction with $W = 1000$~nm, $L = 200$~nm, set the chemical potential to $\mu = 40$~meV and take $\alpha = 50$~meVnm. The choice of such chemical potential guarantees population of the junction with a large number of modes necessary to obtain the Fraunhofer pattern. We perform calculations assuming a zero temperature limit ($T=0$). The critical current maps are obtained via adaptive sampling of the parameter space using Adaptive library \cite{Nijholt2019}. The code used for the calculations is available in an on-line repository \cite{zenodo_repository}. 

\begin{figure}[ht!]
\center
\includegraphics[width = 8cm]{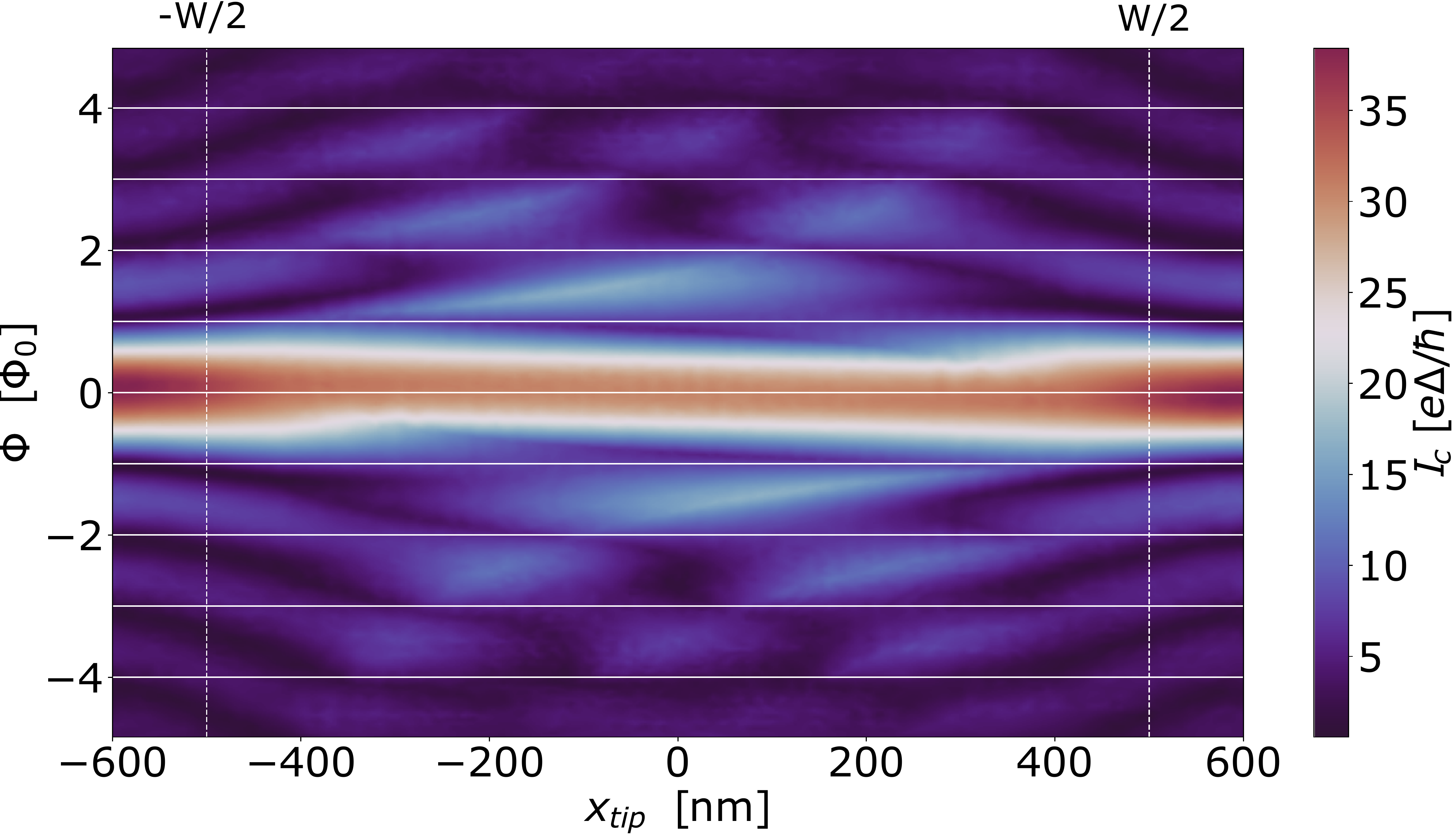}
\caption{Critical current versus flux and tip position (with $y_{\mathrm{tip}} = 0$) obtained in the short-junction limit.}
\label{short_junction_critical_current}
\end{figure}

The critical current versus the flux in the junction $\Phi = BLW$ and the tip coordinate across the junction $x_{\mathrm{tip}}$ (with $y_{\mathrm{tip}} = 0$) is shown in Fig.~\ref{short_junction_critical_current}. When the tip is outside the junction area ($|x_{\mathrm{tip}}| > W/2$) the critical current exhibits the usual Fraunhofer pattern with the periodicity of a single flux quantum $\Phi_0 = h/2e$.  For small magnetic fields, $\Phi < \Phi_0$, for which the critical current without the tip is on the central (first) lobe of the Fraunhofer pattern, $I_c$ decreases as the tip approaches the center of the junction. For a stronger magnetic field, we observe the formation of a pattern of lobes with quenched or increased current emerging from the Fraunhofer pattern as the tip is moved from the edge of the junction to its center. To better understand the physics behind this observation, we now construct a phenomenological model of the SGM $I_c$ response.

\subsection{Phenomenological model of the SGM response}

The critical current of a Josephson junction in a magnetic field can be obtained by tracing the influence of the gauge-invariant phase difference on the supercurrent distribution in the normal part. In a phenomenological model we can thus write \cite{Guiducci2019}
\begin{equation}
I_c = \max_{\phi} \int_{-W/2}^{W/2} J_{S}(x,x_{\mathrm{tip}}) \, \mathcal{I}\left(\gamma(x)\right) dx,
\label{I_c_analytical}
\end{equation}
where the integral is carried out over the junction width. $J_{S}$ is the local supercurrent density per unit length across the junction without the magnetic field, constant along the $y$-direction and suppressed with a Lorentzian function at the tip position  $x_{\mathrm{tip}}$,
\begin{equation}
    J_{S}(x,x_{\mathrm{tip}}) = J_0\left(1-\frac{1}{1+(x - x_{\mathrm{tip}})^2/d^2}\right).
\end{equation}
The supercurrent is modified by the current-phase relationship $\mathcal{I}\left(\gamma(x)\right)$, which can be analytically expressed \cite{PhysRevLett.67.3836} for a junction with transmission coefficient $\tau$ as,
\begin{equation}
    \mathcal{I}\left(\gamma(x)\right) = \frac{\tau \sin(\gamma(x))}{\sqrt{1-\tau\sin^2(\gamma(x) /2)}},
\label{analytical_supercurrent}
\end{equation}
where $\gamma(x)$ is the gauge-invariant local phase shift due to the vector potential $\vec{A}$ and the superconducting phase difference \cite{phasesign},
\begin{equation}
\gamma (x) = \phi + \frac{2 \pi}{\Phi_0} \int_{(x,0)}^{(x,L)} \vec{A} \cdot d\vec{l}.
\label{gamma_equation}
\end{equation}

\begin{figure}[ht!]
\includegraphics[width = 8cm]{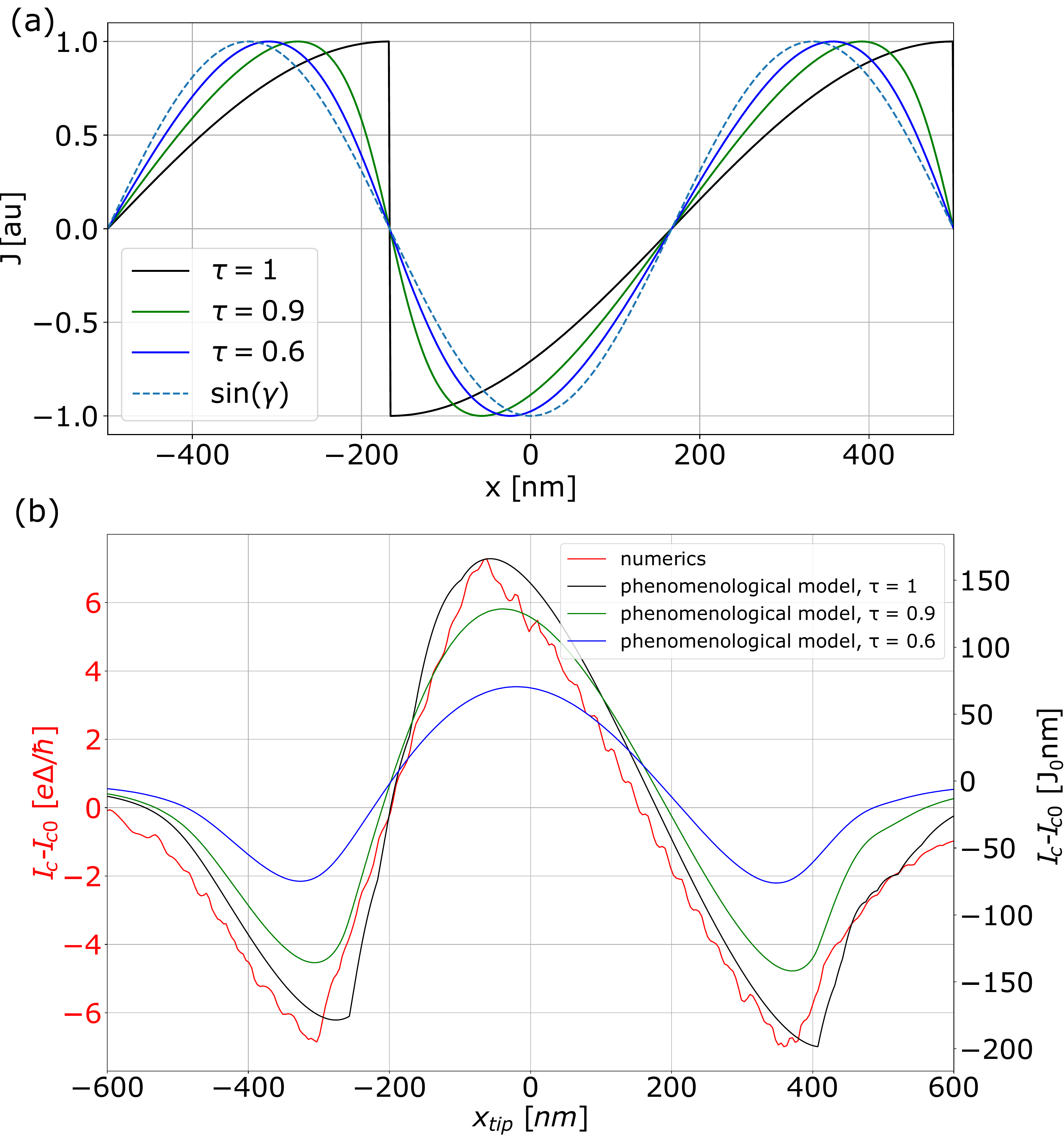}
\caption{(a) Supercurrent distribution across the junction calculated for junctions with different transparency ($\tau$) obtained for constant superconducting phase difference $\phi=-0.5\pi$. (b) Change in critical current obtained in the short-junction calculations (red) and the phenomenological model for various transparencies of the junction. The results are obtained for $\Phi = 1.5\Phi_0$.}
\label{analytics_shift}
\end{figure}

The current-phase relation given by Eq.~\ref{analytical_supercurrent} is skewed for transmittive junctions ($\tau \simeq 1$). This is in contrast to the case of junctions in the tunneling regime ($\tau \ll 1$), where the current-phase relation simplifies to the term $\sin(\gamma)$. Setting the junction transparency to a constant value and inspecting the supercurrent density at given $x$ positions in the junction in a non-zero magnetic field (the phase difference in an external magnetic field depends on the $x$ position through Eq.~(\ref{gamma_equation}) with $\gamma = \phi + 2\pi\Phi x/\Phi_0 W$), we calculate the supercurrent distribution across the junction, as depicted in Fig.~\ref{analytics_shift}(a). We observe that for $\tau = 0.6$ the supercurrent distribution is almost sinusoidal. On the other hand, when the transmission probability is increased, the distribution gets more skewed. 


\begin{figure}[ht!]
\center
\includegraphics[width = 8cm]{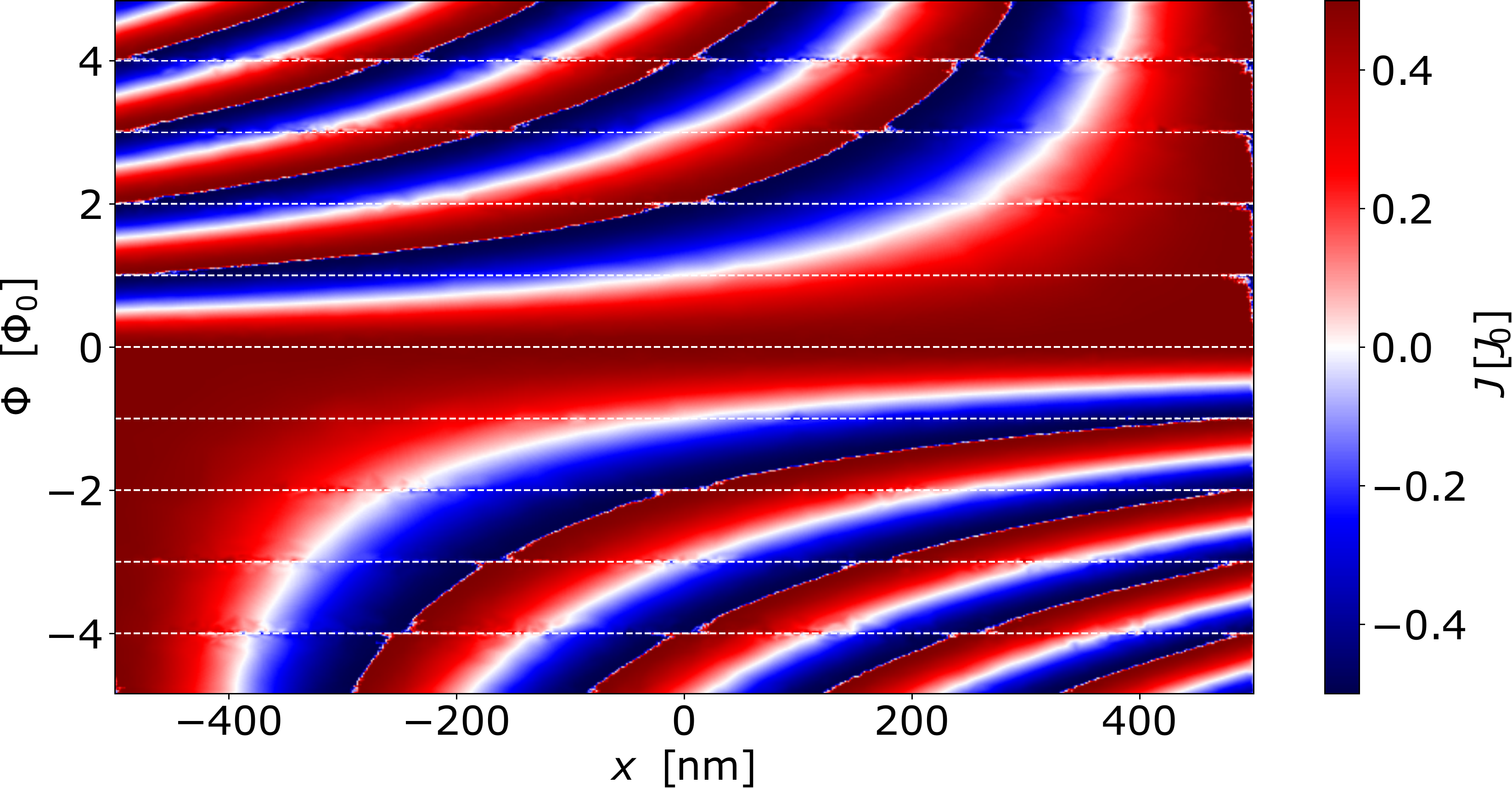}
\caption{Supercurrent distribution $J = J_0 \mathcal{I}(\gamma(x))$ across the junction without the tip obtained in the phenomenological model for $\tau = 1$ versus magnetic field. The current is obtained for the phase difference value $\phi$ that maximizes the supercurrent in the junction.}
\label{analytics_supercurrent_distribution_map}
\end{figure}

The short-junction calculations shown above were obtained for a system in which the only scatterer is constituted by the tip of the SGM. In the phenomenological model, this translates into a high junction transparency. If we consider a ballistic junction with $\tau = 1$, Eq.~(\ref{analytical_supercurrent}) becomes $\mathcal{I}\left(\gamma(x)\right) = \sin(\gamma(x))/4|\cos(\gamma(x)/2)|$. Let us first neglect the tip influence and plot $J = J_0 \mathcal{I}(\gamma(x))$ for $\phi$ set to obtain the maximum supercurrent at a given magnetic field value. The resulting supercurrent distribution across the junction width versus magnetic field is plotted in Fig.~\ref{analytics_supercurrent_distribution_map}. At zero magnetic field ($\Phi = 0$) the supercurrent distribution is constant. When the magnetic field is increased, the vector potential induces a spatially dependent phase shift across the junction, and consequently the supercurrent becomes position--dependent. When $\Phi = n \Phi_0$ (with $n$ an integer), the amount of positive and negative current in the junction is equal. The resulting critical current is zero, and thus a minimum of the Fraunhofer pattern is obtained. An increase of the magnetic field is accompanied by an increasing number of positive and negative supercurrent regions in the junction (as can be seen in the map Fig.~\ref{analytics_supercurrent_distribution_map}) and the number of zero crossings (that corresponds to the Josephson vortices).

\begin{figure}[ht!]
\center
\includegraphics[width = 8cm]{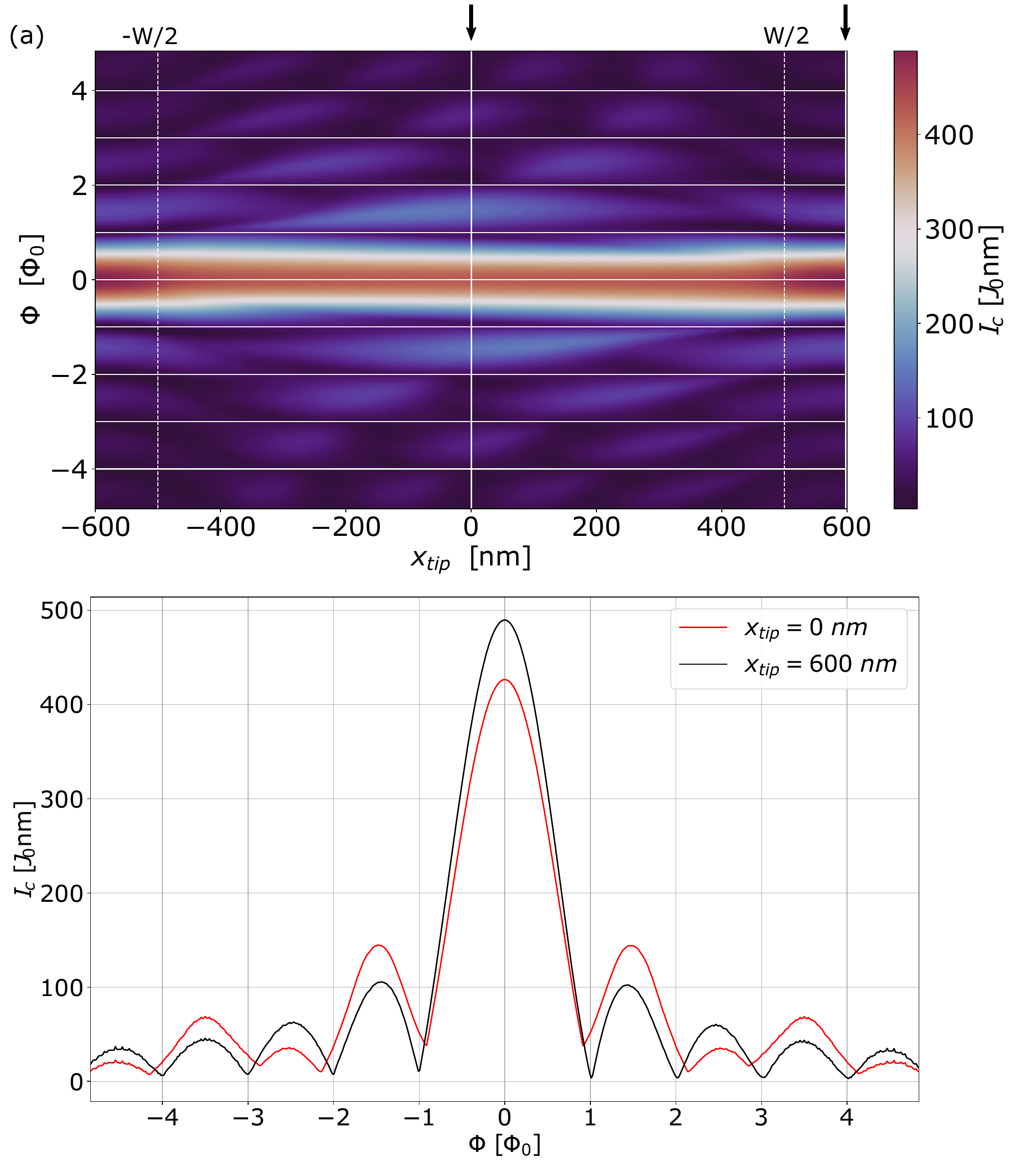}
\caption{(a) Critical current versus position of the SGM tip across the junction and magnetic flux. (b) Critical current cross-sections for two values of the tip position marked with arrows in panel (a).}
\label{analytics_critical_current}
\end{figure}

The critical current obtained in the phenomenological model against the magnetic field and in the presence of the tip is presented in Fig.~\ref{analytics_critical_current}. In panel (a) we see a similar oscillatory pattern as obtained in the short-junction numerical calculation of Fig.~\ref{short_junction_critical_current} (note also the good agreement between the black and red curves shown in Fig.~\ref{analytics_shift}(b)).
In Fig.~\ref{analytics_critical_current}(b) we plot cross-sections of $I_c$ versus the magnetic field for two positions of the tip. When the tip is far from the junction center ($x_{\mathrm{tip}} = 600$~nm), we observe the usual Fraunhofer pattern as obtained from the integration of the supercurrent distribution of Fig.~\ref{analytics_supercurrent_distribution_map} over the $x$ direction. On the other hand, when the tip is located at the center of the junction, i.e.~$x_{\mathrm{tip}}=0$, the height of the lobes is modified depending on their order. An even-odd pattern is obtained, in which odd maxima are suppressed and the even ones are amplified. Furthermore, the periodicity of the positions of the minima with flux quanta is lost.\cite{squidnote}

\begin{figure*}[ht!]
\includegraphics[width = \textwidth]{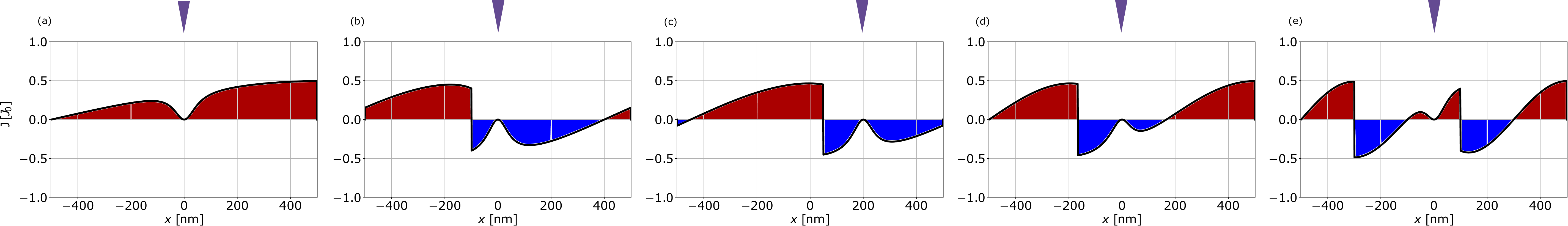}
\caption{Supercurrent distribution across the junction in the presence of the SGM tip, as obtained from Eq.~(\ref{I_c_analytical}). $y_{\mathrm{tip}}=0$. (a) $x_{\mathrm{tip}}=0$~nm, $\Phi = 0.5\Phi_0$, $\phi = 0.5\pi$. (b) $x_{\mathrm{tip}}=0$~nm, $\Phi = \Phi_0$, $\phi= -0.676\pi$. (c) $x_{\mathrm{tip}}=200$~nm, $\Phi = \Phi_0$, $\phi= 0.89\pi$. (d) $x_{\mathrm{tip}}=0$~nm, $\Phi = 1.5\Phi_0$, $\phi= -0.5\pi$. (e) $x_{\mathrm{tip}}=0$~nm, $\Phi = 2.5\Phi_0$, $\phi= 0.504\pi$.}
\label{analytics_supercurrent_vs_tip}
\end{figure*}

To understand how the SGM mapping affects the critical current, let us focus on supercurrent cross-sections obtained in the presence of the tip for four values of the magnetic field, as presented in Fig.~\ref{analytics_supercurrent_vs_tip}.

For a weak magnetic field $\Phi= 0.5 \Phi_0$  [Fig.~\ref{analytics_supercurrent_vs_tip}(a)], we obtain the main supercurrent lobe with an indentation at the position of the tip ($x_{\mathrm{tip}}=0$), as the tip induces a region of density depletion in the junction. Note that the supercurrent value depends on the phase difference $\phi$ in Eq.~(\ref{I_c_analytical}). Its maximal value is obtained for $\phi = 0.5\pi$.

\begin{figure}[ht!]
\includegraphics[width =\columnwidth]{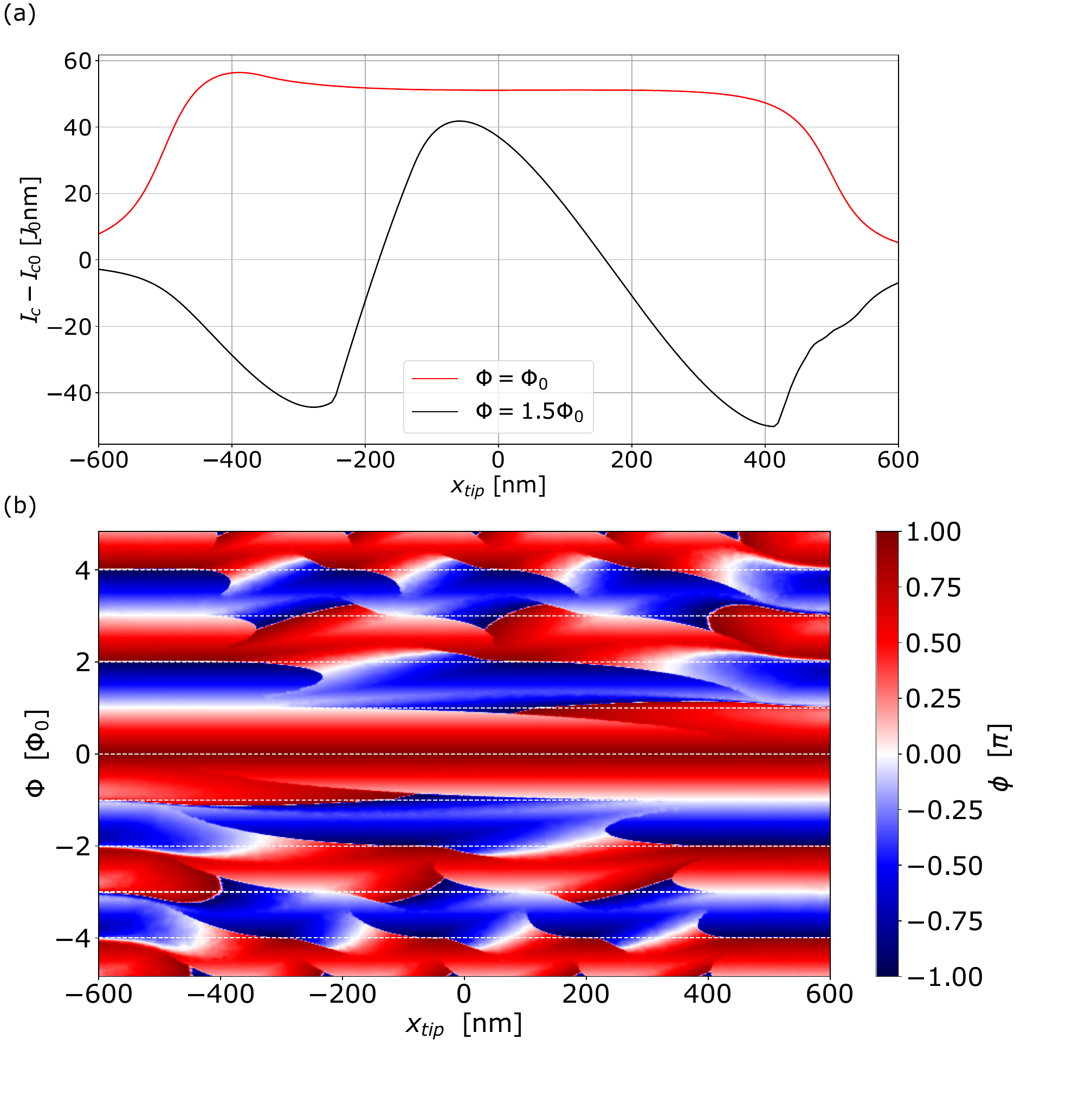}
\caption{(a) Change in critical current versus tip position across the junction with respect to the critical current obtained in the absence of the SGM tip ($I_{c0}$) for two values of the flux as obtained in the phenomenological model. (b) Phase difference that results in maximal supercurrent versus flux and tip position across the junction.}
\label{analytics_phase_changes}
\end{figure}

The supercurrent distribution for the flux $\Phi = \Phi_0$, such that $I_c$ exhibits the first minimum without the tip, is shown in Fig.~\ref{analytics_supercurrent_vs_tip}(b). The amount of positive and negative current is the same, independent of $\phi$, as obtained by considering the magnetic field corresponding to an integer multiple of the flux quantum in Eq.~(\ref{I_c_analytical}), i.e. $\int_{-W/2}^{W/2}\mathcal{I}(\gamma(x))dx = 0$ with $\gamma(x) = \phi+2\pi x n/W$ for integer $n$. When the SGM tip is positioned at the center of the junction, the phase sets to $\phi=-0.676\pi$. Consequently, the tip cuts off a part of the negative current and by that increases the total critical current. Importantly, here the phase is free to adjust to maximize the supercurrent, which results in the negative lobe following the tip position---see Fig.~\ref{analytics_supercurrent_vs_tip}(c) obtained for $x_{\mathrm{tip}} = 200$~nm. As a result, for flux values set to a Fraunhofer minimum, independent on the tip position, the presence of the SGM always increases the critical current---see the red curve in Fig.~\ref{analytics_phase_changes}(a).

On the other hand, at each Fraunhofer maximum, where the number of positive and negative lobes is unequal, there is a unique value of the phase difference $\phi$ that assures the maximal critical current by setting the supercurrent distribution with one extra positive lobe with respect to the number of negative ones. The tip increases $I_c$ if it blocks the negative supercurrent flow, while it decreases the critical current when it suppresses the positive supercurrent lobe---see Figs. \ref{analytics_supercurrent_vs_tip}(d,e). The magnitude of the change of the critical current is proportional to the amount of the suppressed current. This phenomenon links the SGM critical current measurement to the supercurrent distribution across the junction and allows to depict the supercurrent flow [c.f. black curve in Fig.~\ref{analytics_phase_changes}(a) with Fig.~\ref{analytics_supercurrent_vs_tip}(d)] along with its asymmetric distribution. The latter results in breaking of $I_c(x_{\mathrm{tip}}, \Phi) = I_c(-x_{\mathrm{tip}}, \Phi)$ symmetry in the map Fig.~\ref{analytics_critical_current}(a). The features in the map are rather more symmetric with condition $I_c(x_{\mathrm{tip}}, \Phi) = I_c(-x_{\mathrm{tip}}, -\Phi)$ reflecting the skewness of the supercurrent distribution inherent to transparent junctions.

We remark that for junctions with limited transparency, the skewness of the supercurrent distribution is less pronounced, as can be seen in Fig.~\ref{analytics_shift}(a). Accordingly, the asymmetry in the critical current versus tip position is less visible, and the distribution becomes more symmetric when $\tau$ decreases, as can be seen in Fig.~\ref{analytics_shift}(b). Results for opaque junction are given in the appendix.

The aforementioned phenomena allow to connect the regions of strongly amplified (suppressed) critical current in the map of Fig.~\ref{analytics_critical_current}(a) obtained outside of the Fraunhofer minima with the regions of negative (positive) supercurrent distribution in the junction. On the other hand, the process of phase adjustment for flux values close to integer values of the flux quantum is reflected in the strong superconducting phase modulation induced by the tip, visible in Fig.~\ref{analytics_phase_changes}(b).

\subsection{Probing Josephson vortices and supercurrent distribution asymmetry}

\begin{figure*}[ht!]
\includegraphics[width = \textwidth]{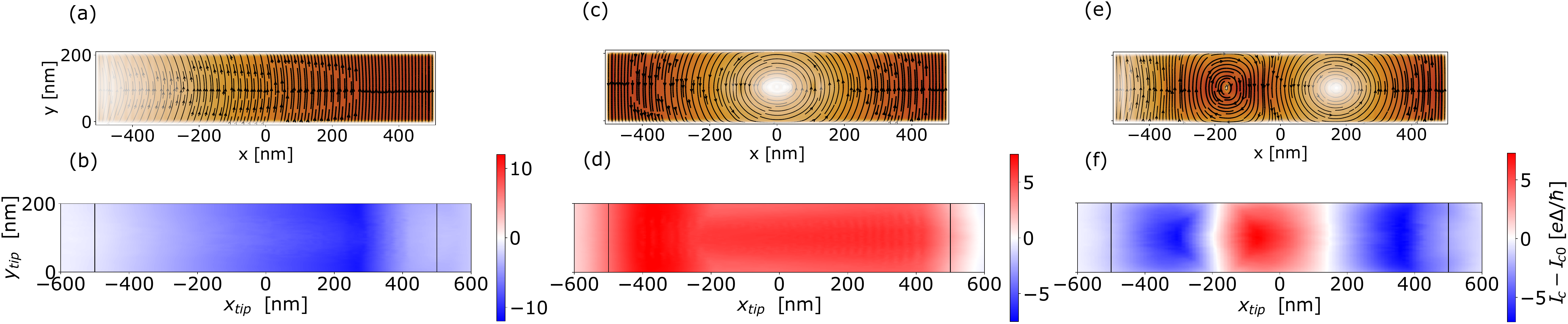}
\caption{(a,c,e) Supercurrent distribution for three values of the magnetic field. (b,d,f) Critical current change  (critical current in the junction minus critical current obtained without the tip) versus SGM tip position. Column (a-b) is obtained for $\Phi=0.5\Phi_0$ with $\phi=0.5\pi$ for (a); Column (c-d): $\Phi=\Phi_0$ with $\phi=0$ for (c); Column (e-f): $\Phi=1.5\Phi_0$ with $\phi=-0.5\pi$ for (e).}
\label{short_junction_current_maps}
\end{figure*}

The supercurrent distribution obtained in the phenomenological model nicely reproduces the features of the supercurrent maps obtained in the short-junction approximation. In Figs. \ref{short_junction_current_maps}(a,c,e) we show supercurrent maps for three values of the magnetic field \cite{current_note}. Panel (a) is calculated for a flux near the first Fraunhofer maximum with $\Phi=0.5\Phi_0$, for the phase difference resulting in the maximal supercurrent. Panels (c) and (e) are obtained at the first Fraunhofer minimum and the second maximum, respectively, where we observe the formation of Josephson vortices.

In an experimentally viable situation, when the critical current is measured in a current-bias configuration, the phase adjusts to provide the maximal supercurrent---at the Fraunhofer minima we always observe a supercurrent increase [Fig.~\ref{short_junction_current_maps}(d)] by the tip. On the other hand, at $I_c$ maxima, the structure of the Josephson vortices is visualized by the SGM [Fig.~\ref{short_junction_current_maps}(f)] in agreement with the results of Sec.~III.A.

Importantly, we observe that in the array of Josephson vortices, the clockwise and anticlockwise vortices have a different spatial span, as can be seen in Fig.~\ref{short_junction_current_maps}(e). In the limit $\tau = 1$, the current abruptly changes sign in the left part of the junction, while on the right it varies smoothly [see the black curve in Fig.~\ref{analytics_shift}(a)]. This is mirrored in the asymmetry of the vortices. The vortex on the left-hand side of the map Fig.~\ref{short_junction_current_maps}(e) is compressed. On the other hand, for positive $x$, the supercurrent increases smoothly as it goes through zero, resulting in an expanded vortex on the right-hand side of Fig.~\ref{short_junction_current_maps}(e). As for the short-junction calculation, the asymmetry in the supercurrent vortices is reflected in the critical current measured, as visible in the map of Fig.~\ref{short_junction_critical_current} as well as in the cross-section presented in Fig.~\ref{analytics_critical_current}(b) with the red curve.

We notice that the asymmetry in counterpropagating vortices has been reported by V. P. Ostroukh et al. \cite{PhysRevB.94.094514} for a square vortex lattice. Here, however, for a linear array of vortices, we demonstrate that the asymmetry bears an important consequence for the distribution of the supercurrent flowing between the superconducting contacts---the current streams are also not symmetric and shifted in the $x$ direction, i.e., the supercurrent flowing in the direction parallel (antiparallel) to the $y$ direction is shifted towards positive (negative) $x$-direction. 

\section{Beyond the short-junction approximation}

The length of the normal part can be of the order of hundreds of nanometers \cite{PhysRevB.93.155402, PhysRevApplied.7.034029, Guiducci2019, PhysRevLett.120.047702, Telesio2021, Ke2019, salimian2021gatecontrolled}, due to fabrication techniques or the requirements of the SGM method itself, i.e., the fact that the SGM tip has to be able to fit between the superconductors, setting a lower bound to the length of the normal part. This combined with a large induced gap when using, e.g., Nb as the superconductor leads to structures that can exceed the short-junction approximation, i.e., $L \simeq \xi$. It is important therefore to assess whether the discussed effects occur also beyond that regime.

To this end, we numerically consider a finite system composed of superconducting leads and a normal, ballistic region (which leads to a creation of a junction with $\tau \simeq 1$). The system has a width $W$ and consist of a scattering region of length $L$ sandwiched between two superconducting regions, each of length $L_{\mathrm{SC}}$. The whole system is described by the Hamiltonian
\begin{equation}
H = 
\left(\begin{array}{cc}
H_N + H_Z\sigma_z& \Delta(x,y)\sigma_0 \\
\Delta^*(x,y)\sigma_0 & -H_N + H_Z\sigma_z\\
\end{array} \right),
\label{long_junction_hamiltonian}
\end{equation}
where $H_Z$ stands for Zeeman term $H_Z = \frac{1}{2}g\mu_BB$. The superconducting pairing potential is
\begin{equation}
\Delta(x,y) = 
\begin{cases}
\Delta & \mbox{if}\;y \le -L/2\\
0 & \mbox{if}\;-L/2 < y < L/2\\
\Delta\exp[i\phi] & \mbox{if}\;y \ge L/2\\
\end{cases}.
\end{equation}
We take $g = -50$, $W = 1000$~nm and $\Delta = 2$ meV. The estimated coherence length is $\xi = 330$~nm and thus shorter than the length of the normal region, which we take as $L=600$~nm. At the same time, it is also shorter than the assumed length of the superconducting leads $L_{\mathrm{SC}} = 400$~nm, allowing the evanescent quasiparticle wave functions to vanish in the superconducting segments. We calculate the energy spectrum of the junction, performing an exact diagonalization of Eq.~(\ref{long_junction_hamiltonian}) discretized on a square lattice. The tip is introduced in a similar manner as in the short-junction calculations, and the supercurrent is calculated from Eq.~(\ref{tot_current}) taking into account the 300 lowest positive energy eigenstates. Zeeman interaction and orbital effects of the field are included only in the normal region.

\begin{figure}[ht!]
\center
\includegraphics[width = 8cm]{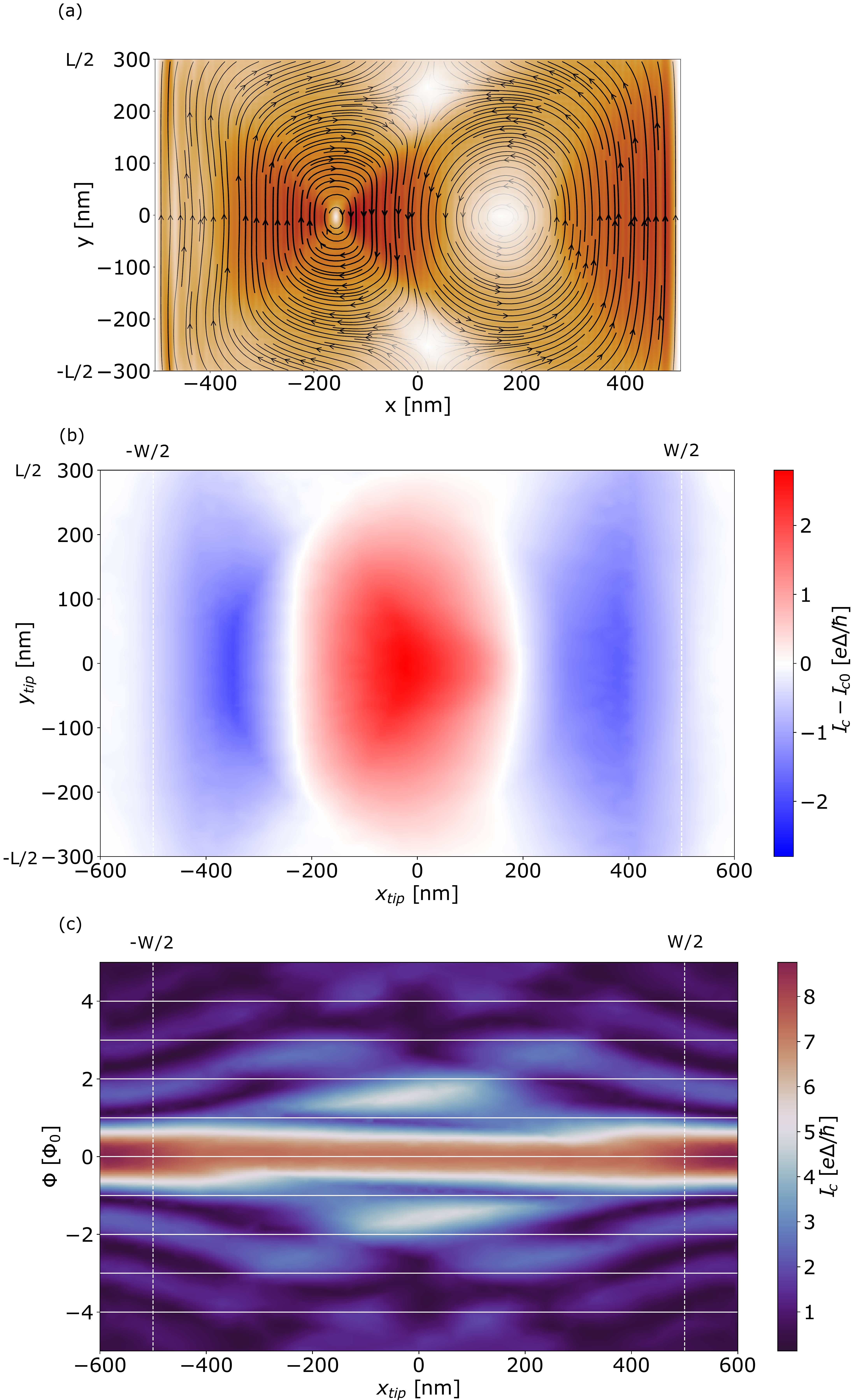}
\caption{(a) Supercurrent distribution for the magnetic field set to the second Fraunhofer maximum and $\phi = -0.5\pi$. (b) Critical current map versus the tip position for the first Fraunhofer maximum. (c) Critical current map versus the tip position across the junction and the flux. The results are obtained in the long-junction regime for $y_{\mathrm{tip}} = 0$.}
\label{long_junction}
\end{figure}

In Fig.~\ref{long_junction}(a) and (b) we show the supercurrent distribution and the corresponding critical current map obtained with the SGM tip for the second Fraunhofer maximum. The supercurrent exhibits similar features as found for the short-junction [Fig.~\ref{short_junction_current_maps}(e)] with a clear anisotropy in the supercurrent and Josephson vortices. Most importantly, for elongated junctions, the vortices form in a linear array at the half length of the junction. The SGM mapping as shown in Fig.~\ref{long_junction}(b) not only determines the supercurrent flow between the superconducting contacts (seen as the change in the critical current for $y_{\mathrm{tip}} = 0$) but also the distribution of the supercurrent along the junction due to the quasiparticles flowing from the leads and circulating around the vortices.

Finally, in Fig.~\ref{long_junction}(c) we show the critical current as a function of the flux in the junction and the horizontal coordinate of the SGM tip (we set $y_{\mathrm{tip}} = 0$). We observe that the Fraunhofer minima are no longer obtained for integer multiples of the flux quantum, as expected for elongated junctions \cite{BARZYKIN1999797}. Most importantly, we obtain a very clear critical current modulation as the tip scans across the sample that can be used to determine the supercurrent distribution together with its asymmetry despite the relaxation of the limits of short-junction approximation.

It should be noted that in very elongated junctions, where the length of the normal region becomes comparable to the Josephson penetration depth, the non-local electrodynamics starts to play a role affecting the supercurrent distribution and also the formation of a Fraunhofer pattern \cite{PhysRevB.55.14486, PhysRevLett.111.117002, PhysRevB.101.144507}. A study of these effects is beyond the scope of this work, since  SNS junctions realized on 2DEGs have typical Josephson penetration lengths of the order of several microns \cite{Barone1982}.

\section{Summary and conclusions}

We theoretically studied scanning gate microscopy imaging of supercurrent flow in planar superconductor-semiconductor-superconductor Josephson junctions. We considered a case where the perpendicular magnetic field induces a spatial variation of the supercurrent and induces Josephson vortices. For systems in both short- and long-junction regimes, we found that the repelling potential of the SGM tip allows to visualize the supercurrent distribution and Josephson vortices via critical current measurements. This is possible provided that the flux in the junction is set outside of Fraunhofer minima for the case without the tip. When the flux is set to a critical current minimum, the superconducting phase difference adapts in such a way that the tip always leads to an increase of the supercurrent. These results are valid both for transmissive ($\tau \simeq 1$) and tunnel ($\tau \ll 1$) SNS junctions. 

Furthermore, for highly transmissive junctions ($\tau \simeq 1$) we found that the scanning gate microscopy critical current maps reveal an asymmetry in the supercurrent distribution, which is caused by the asymmetry between the clockwise and counterclockwise Josephson vortices in transparent junctions. This effect should be already present in recently developed planar junctions where the transmission coefficient takes high values of $\tau \simeq 0.9$ \cite{PhysRevApplied.7.034029, salimian2021gatecontrolled}.

Our results were obtained assuming a short-junction regime and supported by a phenomenological model, where the tip suppresses a phase-dependent current in the junction. Extension beyond the short-junction limit was implemented to confirm the robustness of our findings.

\section{Acknowledgement}

We acknowledge helpful discussions with Alina Mreńca-Kolasińska. This work was supported by National Science Centre (NCN) agreement number UMO-2020/38/E/ST3/00418. SH and MC were partially supported by the SUPERTOP project, QUANTERA ERA-NET Cofound in Quantum Technologies (H2020 Grant No. 731473) and by the FET-OPEN project AndQC (H2020 Grant No. 828948).

\appendix*
\section{Phenomenological model in the tunneling limit}
\label{Appendix}
In the tunneling limit, when $\tau$ is much smaller than 1, the current phase relation Eq. (\ref{analytical_supercurrent}) can be approximated by $\mathcal{I} = \sin(\gamma(x))$. This results in a symmetric supercurrent distribution in the junction as shown in Fig.~\ref{analytics_shift}(a) with the green-dashed curve. Here we provide results for the tunneling case that complement the phenomenological analysis done in the main text.

\begin{figure}[ht!]
\center
\includegraphics[width = 9cm]{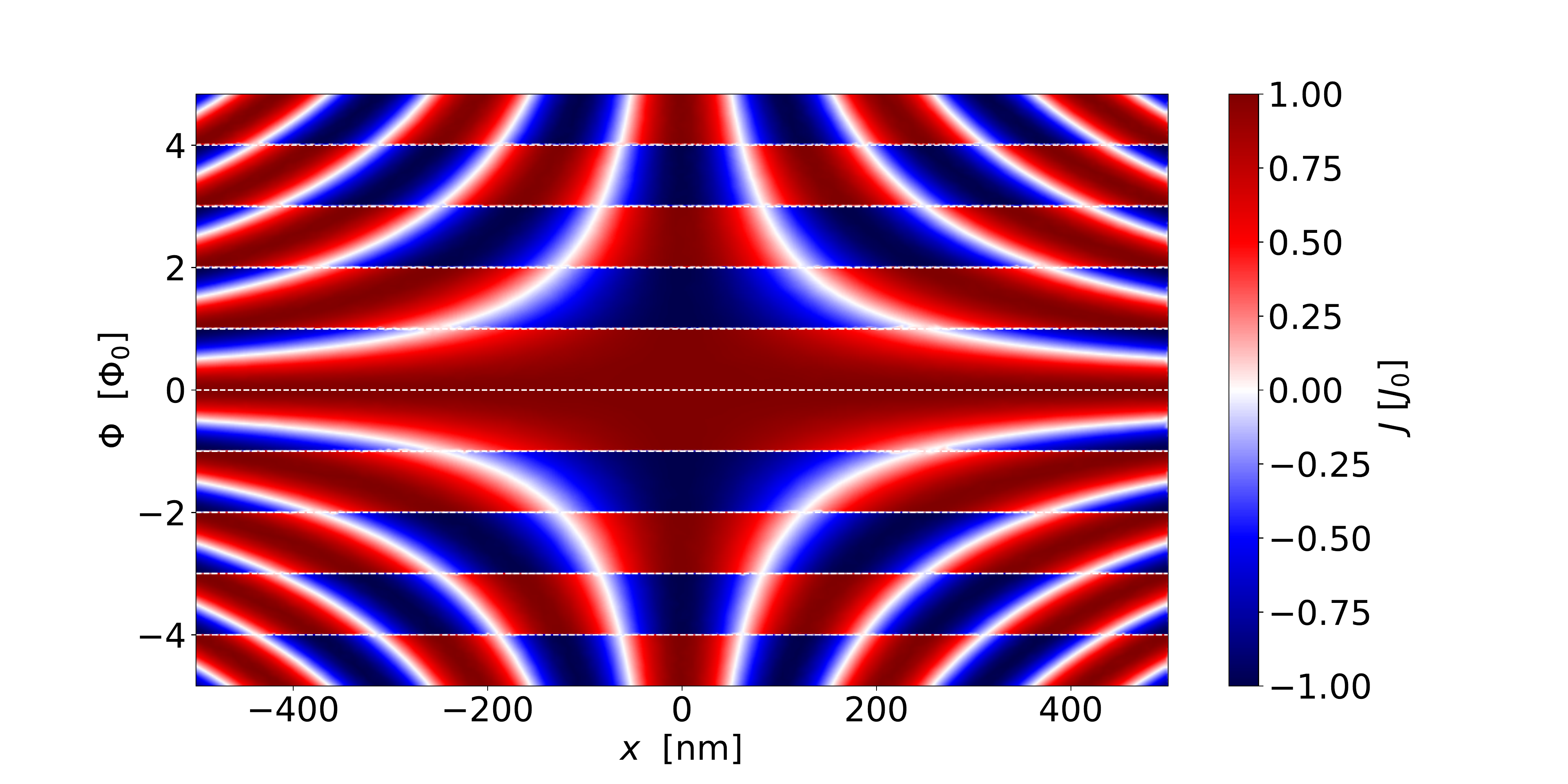}
\caption{Supercurrent distribution $J = J_0 \mathcal{I}(\gamma(x))$ across the junction without the tip obtained in the phenomenological model for $\mathcal{I} = \sin(\gamma(x))$  versus magnetic field. The current is obtained for the phase difference value $\phi$ that maximizes the supercurrent in the junction.}
\label{supplementary_supercurrent_distribution}
\end{figure}

\begin{figure}[ht!]
\center
\includegraphics[width = 8cm]{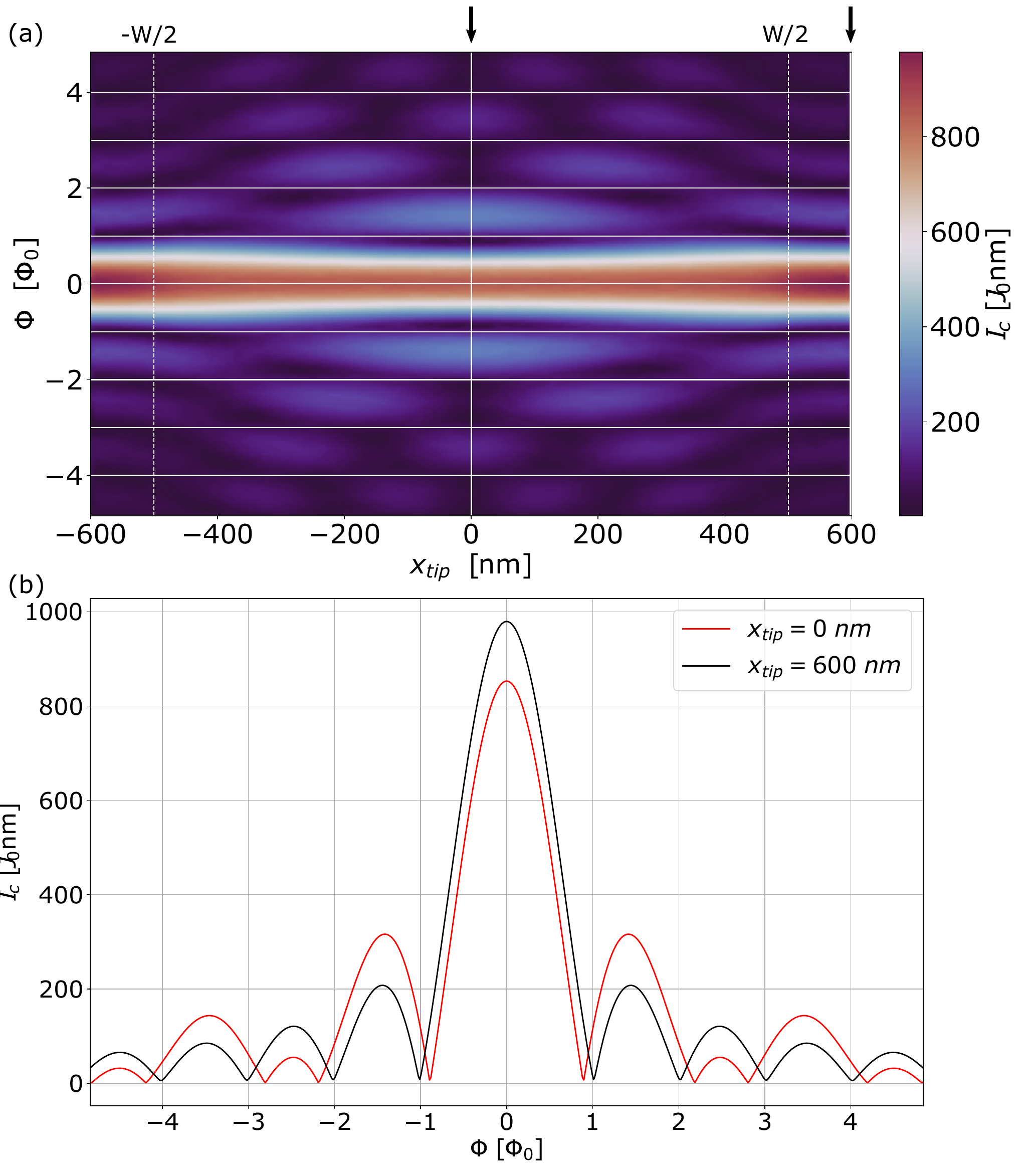}
\caption{(a) Critical current versus magnetic flux and position of the SGM tip across the junction. (b) Critical current cross-sections for two values of the tip position marked with arrows in panel (a).}
\label{supplementary_critical_current_map}
\end{figure}

The supercurrent distribution presented in Fig.~\ref{supplementary_supercurrent_distribution} shows formation of positive and negative supercurrent regions. When $\Phi = n\Phi_0$ (with $n$ an integer), the amount of positive and negative current in the junction is equal—a complete cycle of supercurrent is formed. The resulting critical current is zero, and so a minimum of the Fraunhofer pattern is obtained. Every time the complete cycle is crossed, the phase jumps by $\pi$ and the critical current increases until it reaches the next Fraunhofer maximum---see Fig.~\ref{supplementary_critical_current_map}.

\begin{figure*}[ht!]
\includegraphics[width = \textwidth]{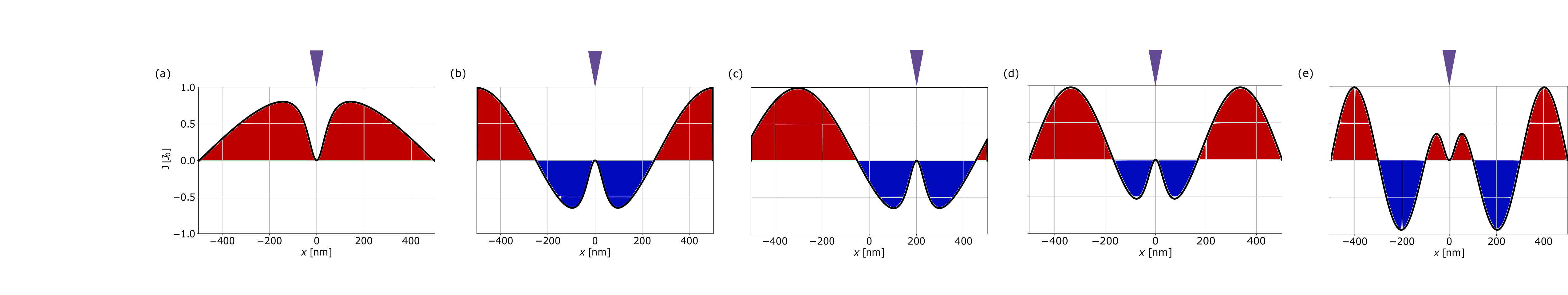}
\caption{Supercurrent distribution across the junction in the presence of the SGM tip at $y_{\mathrm{tip}}=0$. (a) $x_{\mathrm{tip}}=0$~nm, $\Phi = 0.5\Phi_0$, $\phi = 0.5\pi$. (b) $x_{\mathrm{tip}}=0$~nm, $\Phi = \Phi_0$, $\phi= -0.5\pi$. (c) $x_{\mathrm{tip}}=200$~nm, $\Phi = \Phi_0$, $\phi= 0.9\pi$. (d) $x_{\mathrm{tip}}=0$~nm, $\Phi = 1.5\Phi_0$, $\phi= -0.5\pi$. (e) $x_{\mathrm{tip}}=0$~nm, $\Phi = 2.5\Phi_0$, $\phi= 0.51\pi$.}
\label{supplementary_supercurrent_vs_tip}
\end{figure*}

The symmetry in the current distribution shown in Fig.~\ref{supplementary_supercurrent_distribution} translates into symmetric features in the SGM critical current map shown in Fig.~\ref{supplementary_critical_current_map}.

The process of critical current modification operates in the same manner as for a transparent junction (cf.~Fig.~\ref{analytics_supercurrent_vs_tip}). At Fraunhofer minima it is always the negative supercurrent region that is suppressed [Fig.~\ref{supplementary_supercurrent_vs_tip}(b,c)], which results in an increase of $I_c$ by the tip. At Fraunhofer maxima [Fig.~\ref{supplementary_supercurrent_vs_tip}(d,e)] the tip suppresses the supercurrent in its vicinity decreasing or increasing the critical current, depending on the sign of the suppressed supercurrent. Importantly, as now the supercurrent distribution is symmetric, so are the critical current features, as show in the map Fig.~\ref{supplementary_critical_current_map}(a).

\bibliography{references.bib}

\begin{thebibliography}{65}
\bibitem [{sho()}]{shortjunctionnote}%
  \BibitemOpen
  \href@noop {} {}\bibinfo {note} {In the literature there is another meaning
  of "short-junction" which refers to the junction shorter than the Josephson
  penetration depth, where the self field effects can be neglected. In this
  sense all the junctions considered in this work belong to this
  regime.}\BibitemShut {Stop}%
\bibitem [{pha()}]{phasesign}%
  \BibitemOpen
  \href@noop {} {}\bibinfo {note} {Note that the gauge invariant phase results
  from a gauge invariant phase gradient $\nabla \gamma = \nabla \phi - q/\hbar
  \mathbf{A}$ which leads to a plus sign in the last term of
  Eq.~(\ref{gamma_equation}) when using nomenclature $\Phi_0 = h/2e >
  0$.}\BibitemShut {Stop}%
\bibitem [{squ()}]{squidnote}%
  \BibitemOpen
  \href@noop {} {}\bibinfo {note} {By properly tuning tip height and applied
  bias, so that a large area of the junction becomes depleted, the SGM could be
  used to continuously tune the junction from a Fraunhofer pattern to a
  SQUID-like pattern.}\BibitemShut {Stop}%
\bibitem [{cur()}]{current_note}%
  \BibitemOpen
  \href@noop {} {}\bibinfo {note} {The wave functions in the normal region
  obtained from the solution of the scattering problem defined by
  Eq.~(\ref{HN}) are phase independent. On the other hand, the ABS
  wave-function amplitudes (obtained as eigenvectors of
  Eq.~(\ref{ABS_eigenequation})) are orthonormalized---as a result the
  information about the magnitude of the current carried by each ABS is not
  contained in its wave function. For a proper calculation of the supercurrent
  distribution one needs to normalize the supercurrent carried by each ABS
  accordingly to its value obtained directly from the ABS energy
  spectra.}\BibitemShut {Stop}%
\end{thebibliography}

\end{document}